# PROGRESS ON THE ISIS SYNCHROTRON DIGITAL LOW LEVEL RF SYSTEM UPGRADE

A. Seville[†], D. B. Allen, I. S. K. Gardner, R. J. Mathieson, STFC, Didcot, UK

*Abstract*

The ISIS synchrotron at the Rutherford Appleton Laboratory in the UK now routinely uses a dual harmonic RF system to accelerate beam currents in excess of 230 µA to run two target stations simultaneously. The acceleration in the ISIS synchrotron is provided by six fundamental frequency (1RF) and four second harmonic (2RF) RF cavities. The 1RF systems are required to sweep from 1.3MHz to 3.1MHz during the 10ms acceleration period, repeated at 50Hz, with the 2RF systems sweeping from 2.6MHz to 6.3MHz. The existing analogue LLRF control system has been in service for over 30 years and is now showing some signs of old age and spare parts are becoming difficult to source. In order to overcome this and to give more stable control of the phase of the RF voltage at each of the cavities, changes have been made to the LLRF control system. A new FPGA based combined frequency law generator / master oscillator has been implemented using "off-the-shelf" National Instruments PXI-express based FlexRIO modules. This initial design has been successfully used during the ISIS operational cycles for over three years. This paper reports on the commissioning of the FlexRIO system, the implementation and recent testing of the cavity control loops and plans for the gradual replacement of remaining parts of the LPRF system.

## INTRODUCTION

The ISIS synchrotron supplied first beam to the target in December 1984. 70MeV H⁻ ions are injected into the ring where they are stripped and accelerated to 800MeV before extraction and transport to the target station (TS1).

The proton beam is accelerated by six, H=2, ferrite loaded RF cavities. These RF cavities are biased with a current ranging from 200A to 2000A during the 10ms accelerating period to tune the cavity from 1.3MHz at injection to 3.1MHz at extraction. Since then, there have been several upgrades to the machine. These include the addition, at the turn of the millennium, of four second harmonic (H=4) RF cavities [1], sweeping from 2.6MHz to 6.3MHz. These enabled the stable acceleration of the higher beam current needed to supply a second target station (TS2 at 10Hz) and whilst maintaining the same mean beam current to TS1 and the intermediate muon target at 40Hz.

The ISIS Synchrotron LLRF system provides the RF signal to each of the 10 accelerating cavities. The RF sweep is produced by the Frequency Law Generator (FLG) which takes in a $\dot{B}$ signal from a search coil within one of the Synchrotron dipole magnets. This is integrated to give B and then mapped to an output voltage corresponding to the RF frequency. This voltage is then summed with the outputs of the 3 beam loops: the beam phase loop (which damps beam dipole oscillations), the radial loop (which maintains the horizontal position of the beam) and the bunch length loop (designed to damp quadrupole beam oscillations and ease extraction between the two bunches) and then fed into the Master Oscillator.

Within the beam loops, the RF system has individual control loops to control the amplitude, phase and tuning of each RF cavity as depicted in figure 1.

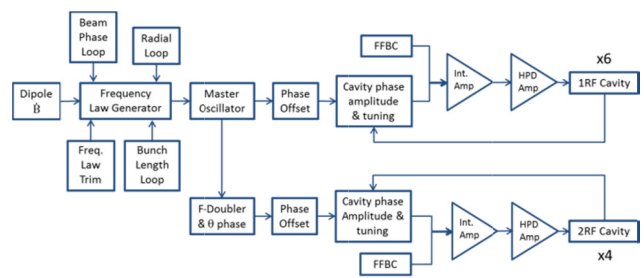

Figure 1: ISIS Synchrotron LLRF Control System.

The original analogue controls for these systems have been in place, now, for over 35 years. Even the more recent 2RF controls are showing signs of ageing and sourcing replacement components is becoming increasingly difficult.

## INITIAL OBSOLESCENCE PLANNING

Following the commissioning of the 2RF systems in 2004, plans were made to design replacement units for the FLG and Master Oscillator [2] modules. These were based on the then new Lattice FPGA devices. These modules were never fully commissioned, partly due to staff turnover, but their designs were used to inform the later Digital-LPRF upgrade project.

At this time, plans were made to upgrade the feed-forward beam compensation (FFBC) units. These had been installed to compensate for beam loading of the RF cavities as beam intensities increased by applying a variable gain and delay to a beam intensity signal, before summing with the RF drive signal to the cavity from the LLRF system. The upgrade design included a swept digital filter implemented on the new FlexRIO FPGA module, within a National Instruments PXI crate.

It soon became apparent that the FlexRIO platform might be able to provide a means of replacement for more of the LLRF control system. To this end an initial Master Oscillator design was implemented on the FPGA module to investigate the potential for this approach. The design was produced in around six months, which compared favourably to the 3 years it had taken for the bespoke Lattice FPGA - based design to be built and configured.

___
† Andrew.Seville@stfc.ac.uk

The use of the FlexRIO platform fitted in well with our limited resources: a small RF team already supporting operation of an operational machine running 24/7 for 200 days per year. Buying "off-the-shelf" hardware should save considerable development time. STFC have a LabView licence agreement in place with National Instruments, which includes use of the LabView FPGA development module, so there was a strong LabView user community on site at RAL, with much experience to draw on both for support and for future succession planning.

## DIGITAL LLRF DEVELOPMENT

A staged implementation of the digital LLRF system was adopted, starting with the replacement of the Master Oscillator (MO), and then expanding the digital system to replace more and more of the existing system. This was suited to the reconfigurable nature of the FPGA devices and would also allow confidence in the new system to grow following some reliability issues with previous projects using Windows PXI controllers within our Beam Diagnostics section. The process of reverting back from the deployed digital system to the analogue LLRF system would have to be simple and fast, if necessary. The initial FPGA code was designed offline throughout the first period of development. Testing the system live with beam, however, was limited to the machine physics shifts usually scheduled within a 2-3 day period at the end of each user cycle.

The MO code was implemented on a single FPGA module and front end transceiver adapter in December, 2013. In August 2014, this had been expanded to include the functionality of the beam loop summing amplifier and Master Oscillator, within two FPGA modules in the PXI crate, as shown in figure 2.

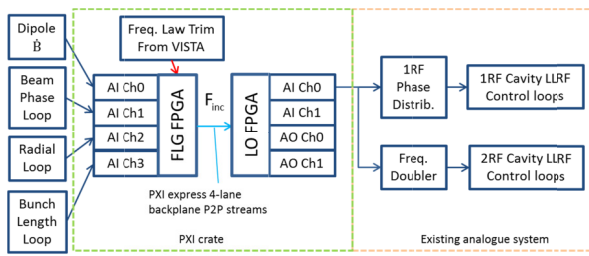

Figure 2: Digital FLG /Master Oscillator Configuration.

The $\dot{B}$ and Beam loop signals were fed into a NI5734 4-channel digitiser adapter, attached to the FLG FPGA module, on which a 17 bit word was generated corresponding to the frequency increment ($F_{inc}$). $F_{inc}$ was then transmitted across the PXI-express backplane via Peer-to-Peer (P2P) streaming to a Local Oscillator (LO) FPGA, and used as the counter increment in a DDS, which was then output on one of the DAC channels of the NI5781 Transceiver adapter and fed into the existing analogue system in place of the original Master oscillator.

A second LO system was then added which included a frequency doubler to generate the frequency sweep, with the additional dynamic phase offset (θ) required for the 2RF systems. The LO transceiver adapters were upgraded to a NI5782 model with two input/output channels sampled at up to 250MHz. This was successfully tested with beam in August 2014.

However, when the above system was expanded to be used with more LO FPGAs, the increased time taken to send the $F_{inc}$ word over multiple P2P streams, combined with the PXI-express switch fabric delay, required the $F_{inc}$ update period to be increased from 2.6μs to 5.2μs. This delay was compounded by the use of the PXI-express backplane for system configuration data, which sporadically disrupted the broadcast data stream. This manifested itself during beam tests in instability in the beam phase loop leading to dramatic beam losses.

A faster means of broadcasting $F_{inc}$ was implemented by modulating the 17 bit word over the four of the PXI-e trigger lines, with a 40MHz bit-rate (on each line), which was successfully tested with beam in April 2015.

In February 2016, the system was used to supply the RF signals to the analogue LLRF system throughout the ISIS user cycle. The system was configured to provide a single 1RF input sweep to the analogue LLRF system and a θ phase modulated 2RF sweep into each of the 2RF analogue cavity control loops as shown in Figure 3.

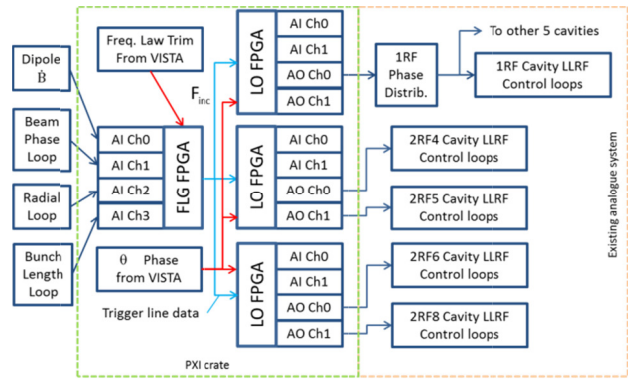

Figure 3: D-LLRF Configuration for first user cycle.

The Frequency law trim function and the θ phase function were obtained from the ISIS VISTA controls via a UDP based LabView program, dubbed "PixyBroker", developed by Tim Gray in the ISIS Controls Group.

This system, with occasional code updates, formed the basis of ISIS operations for the next three years, during which work to develop the cavity voltage control loops continued.

## CAVITY TUNING

The analogue RF cavity tuning loop is controlled by minimising the phase error between the cavity gap Voltage monitor and a monitor of the Tetrode grid voltage. The cavity tuning loop bandwidth is 5kHz or so, and so the loop error is reduced by applying a calculated "Cav-tune" correction function [3] before sending the signal to the cavity bias regulator, as shown in figure 4.

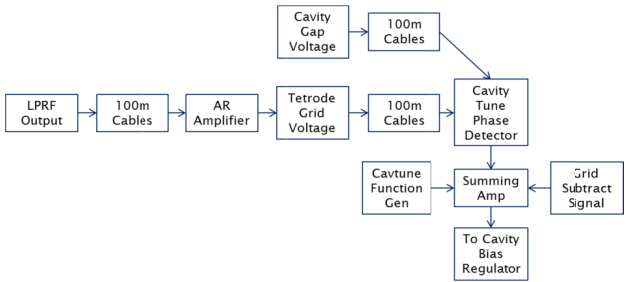

Figure 4: RF cavity tuning.

For the 2RF systems, under heavy beam loading, the level control loop reduces the grid voltage down towards 0V, as the required gap Voltage is supplied by the beam itself. The cavity tuning phase detector then cannot operate and the tuning loop becomes unstable, as shown in figure 5.

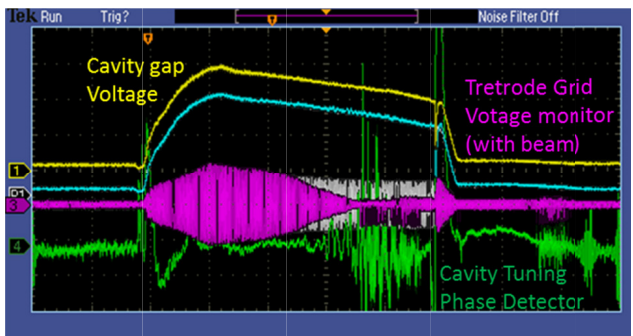

Figure 5: Tuning loop signals under heavy beam loading.

Assuming that the system response from the output of the LLRF system down the cables to the Tetrode grid and back up the monitor cables to the phase detector is a simple delay, the grid Voltage signal into the cavity tuning phase detector can be replaced by a fixed amplitude reference. This approach was implemented and used operationally for the analogue 2RF systems for the last three years or so.

## DEPLOYED SYSTEM ARCHITECTURE

The D-LLRF system as currently deployed is shown schematically in figure 6.

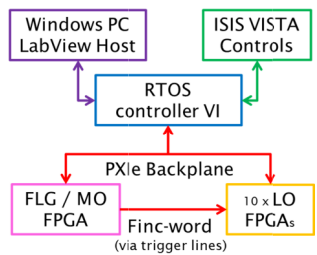

Figure 6: D-LLRF System Architecture.

The NI-8135 RealTime Controller operates the "Boot-up" executable LabView Virtual Instrument (VI) that downloads the bitfiles to each FPGA on power up, performs initialisation and clock synchronisation of the FPGA Modules and then marshals data from the Host VI program and ISIS VISTA controls system to the FLG and each of the ten LO FPGA Modules. The PixyBroker VI is currently still used to pass data to and from the controls system, where parameters are set by machine operators. This relies on polling of the VISTA database channels and a faster MQTT based method is currently being tested which will make parameter updates event driven.

The Windows Host VI provides a local user interface with tabbed panes to access the RF setup parameters (usually only accessed by RF experts). These include an FLG setup pane which contains Beam loop gain settings and displays the top and bottom frequency for the RF sweep. The status tab includes deployment status for the LLRF system and a simple LPRF On/Off button. The LO FPGA tab allows interactive editing of the setup parameters for each cavity, such as phase offsets, Open / Closed loop operation and PI Loop gains. A further tab displays the virtual "Function Module" channels with "last sent" values of function profiles from the VISTA controls system, eg the demanded gapvolts amplitude profile

The Windows PC is also used to operate a virtual oscilloscope VI, which can update up to 4 simultaneous, 10000 point channels at up to a 50Hz refresh rate, as shown in figure 7. The 4 virtual signals are generated and selected in each FPGA and streamed over the PCI-express backplane to the RT controller where the channels for a single FPGA can be selected and sent via a network stream for display on the host PC.

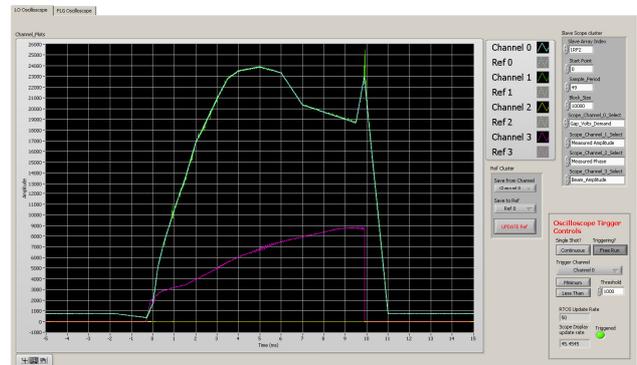

Figure 7: Host Oscilloscope VI.

Triggering of the display is currently applied by thresholding the displayed signals on the PC VI, but local triggering on each VI will soon be implemented which should reduce the traffic on the PXI-e bus.

The FLG FPGA code is implemented using LabView FPGA on the NI-7966R FPGA module and NI-5734 120MS/s, 4 channel, 14 bit digitiser module. The code is shown schematically in figure 8.

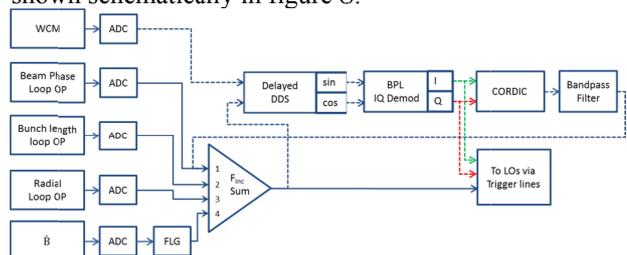

Figure 8: FLG FPGA System Diagram.

The $\dot{B}$ signal, together with the 3 analogue beam loop outputs are digitised on the front end adapter and passed into the FLG FPGA. Here, the $\dot{B}$ signal is integrated to give B, which is then mapped to $F_{inc}$ via a look up table. The $F_{inc}$ word is summed with a scaled version of the Frequency law Trim function obtained from the VISTA controls system and scaled versions of the digitised loop signals, before being transferred to the local oscillator FPGA modules by the trigger line modulation scheme.

Previous tests successfully used *IQ* demodulation of a digitised beam sum electrode signal followed by a CORDIC algorithm to generate a beam phase signal. Figure 9 shows a comparison of the beam phase signal measured using this approach with that generated by the existing analogue beam phase loop. This will soon be implemented on the operational system to replace the existing analogue beam phase loop. The same beam signal may then be used to generate the Bunch Length Loop correction and possibly provide Beam *I* and *Q* components for use in Feed Forward Beam Compensation.

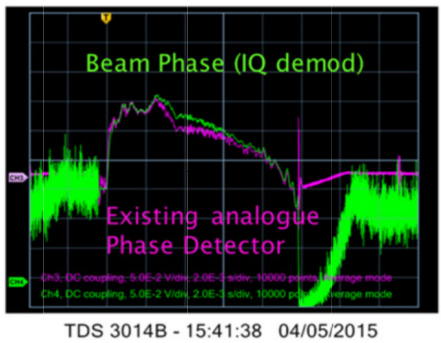

Figure 9: Analogue and Digital Beam Phase Detection.

The FLG FPGA module is also used to distribute the Frame start trigger and machine timing signals to the other FPGA modules, which will enable pulsing of experimental parameter values at repetition frequencies ranging from 50Hz down to $50/640^{th}$ Hz. This will also allow for triggering different parameter values for TS1 beam and TS2 beam.

The LO FPGA code, shown schematically in Figure 10, is implemented on each of the LO NI-7966R FPGA modules.

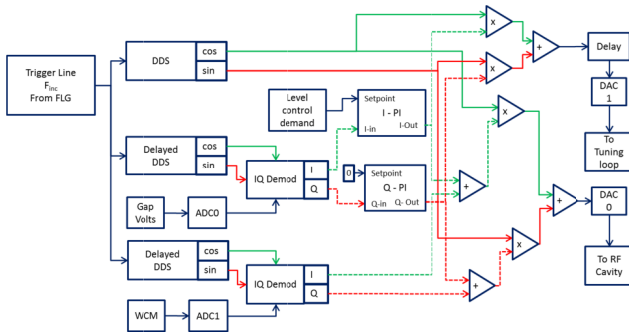

Figure 10: LO FPGA System Diagram.

The Gap Voltage monitor signal is digitised on the NI-5782 transceiver adapter at 250MS/s and passed to the LO FPGA. The signal is decimated, appropriately scaled and passed into an IQ-demodulator block along with a delayed reference signal generated from the received $F_{inc}$ word used to step through a DDS look up table. The *I* and *Q* components are each used as process variables in separate PI loops. The setpoint for the *Q* loop is set to zero and that for the *I* loop is a scaled version of the amplitude demand profile obtained from the VISTA control system.

The beam wall current monitor signal is also digitised on the $2^{nd}$ analogue input channel of the transceiver adapter module. This is then *IQ* demodulated to give $I_{beam}$ and $Q_{beam}$ correction components. These are then added to each of the *I* and *Q* PI loop outputs and the result used to feed the *IQ*-modulation of a DDS RF sweep, which is in turn fed to the RF cavity via a transceiver DAC channel.

The *I* and *Q* PI loop outputs above are used directly to modulate an RF sweep which is then passed through a pipeline delay and output on the $2^{nd}$ DAC channel of the transceiver adapter. This can then be fed directly into the analogue cavity tuning phase detector in place of the Tetrode grid Voltage.

## *IQ* LOOP OPERATION

The system architecture described above was commissioned to replace the analogue amplitude & phase control loops on a single 1RF cavity during machine physics shifts in February and Ma rch 2019. This system was used to accelerate a high intensity beam (~230μA) at 40Hz to TS1, with low beam losses. No feed forward beam compensation was applied to the digitally controlled cavity during these tests. Operation at this intensity using the analogue system without feed forward beam compensation is not possible, as the beam loading can cause the level control loop to output a very small amplitude signal, causing the cavity phase and tuning loops to become unstable. The *IQ*-loop gives better performance even for the low amplitude case, as the *I* and *Q* components can be driven negative, but the error functions for both loops signal remain stable.

Further tests on a single *IQ* loop controlled cavity were carried out with both the delayed output tuning and feed forward beam compensation applied in the LO FPGA. The measured amplitude and phase for this cavity during beam injection and for the first 3ms of acceleration are shown in figure 11, compared with those for a cavity controlled by the old analogue system.

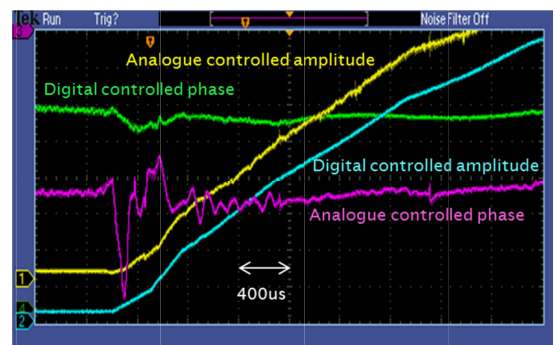

Figure 11: Analogue and Digital Controlled.

The phase transient during beam injection for the analogue amplitude / phase controlled cavity was ~30º compared with ~5º for the digital *IQ* controlled cavity. One can also see that the *IQ* controlled cavity amplitude profile appears much smoother.

These results were very encouraging and plans were made to deploy the design for all ten RF cavities. Several more FPGA modules would be required for the operational deployment for this number of systems, with sufficient spares. In early 2019, the FLG FPGA module and 4-channel digitiser were replaced by a Kintex 7 PXIe7971R FPGA module and NI5783 4-channel transceiver adapter, as the FPGA utilisation was approaching 90% and a larger device would speed up compilation times and the additional 4 DAC channels on the new transceiver adapter could then be used to generate auxiliary signals such as the Frequency Law signal sent to the Beam Intensity monitor and an RF sweep used for machine extract timing. This would also allow the PXIe7966 FPGA modules already purchased as FLG FPGAs to be re-deployed as spare LO FPGA modules.

The new design with a digital *IQ* controller was deployed on the single 1RF cavity for the beginning of the ISIS user cycle in June 2019. However, 3 hours into the user cycle, the system lost synchronisation between some of the LO FPGAs, causing dramatic beam losses. Previously the system had been running stably for more than a week. The 4 channel digitiser adapter and Virtex-5 FPGA module were reinstated as FLG FPGA and beam was restored for the rest of the user cycle.

Subsequent investigations found differences in the clock implementation on new 7971R FPGA module leading to additional 2.5ns delay. This was just sufficient to cause sporadic loss of synchronisation. The FLG FPGA code was then changed to reduce the $F_{inc}$ bit clock rate to 20MHz (previously 40MHz), which would give sufficient room for any jitter on the bit transfer to fall well within the $F_{inc}$ bit update clock period.

The updated FLG code was commissioned on the Kintex-7 FPGA module and 4 channel transceiver adapter and this was deployed for the beginning of the ISIS user cycle starting on 10th September and has been successfully running to date. The six fundamental cavities are now controlled with the digital *IQ* loop, though due to problems elsewhere on the machine, available time to setup the digital feedforward beam compensation the systems was limited, so we are operating with traditional grid-voltage input to the cavity tuning loop and also using the analogue FFBC system until the next user cycle.

## CONCLUSIONS & FUTURE WORK

The choice of the NI PXIe platform gave a fast route to implement the initial simple system designs but did then add constraints to the full design, requiring some lateral-thinking solutions to be developed. The reconfigurable aspect of the FPGA modules enabled a gradual implementation of the system, which built confidence during the initial operation as a combined Frequency law generator / Master oscillator, and gave the ability to test individual 'add-on' pieces of code functionality during machine development time, prior to operational deployment. This approach may have led to a longer development time than designing a new, fully functioning system.

However, the system has now been successfully commissioned for operational use in controlling the ISIS synchrotron RF cavities. The use of a digital *IQ* loop has given improved performance over the existing analogue amplitude and phase loops. Should feedforward beam compensation still be necessary, it will be implemented within the FPGA, allowing the ageing analogue beam compensation units to be removed and reduce machine downtime.

The digital *IQ* loop control will be deployed for all 10 RF cavities, with triggered virtual oscilloscope signals and an updated ISIS VISTA controls interface in early 2020. Also, the FLG FPGA code will be updated to use the Wall Current monitor signal to directly operate the beam phase and also the bunch length loops and also to provide the auxiliary signals for extraction triggering etc. Further development is required to operate the cavity tuning loop (beyond the current production of the RF sweep fed into the existing analogue system). This may require moving the feed forward beam compensation code onto the FLG FPGA, to allow room for the cavity tuning *IQ* demodulation on the LO FPGAs.

Future work will include updating the LLRF system in response to changing requirements caused by upgrades to the cavity high power drive amplifiers and bias regulators. We also aim to reduce the RF power budget by 10-20% with the use of beam triggered RF.

The LLRF team will be providing support for the LLRF system for the ISIS Front Test Stand [4,5], which has an initial design using the same NI PXI-e platform hardware. Many of the techniques and code (eg VISTA controls interface, IQ loops, etc.) may be re-used for this project.